\documentclass[12pt, letterpaper, oneside]{article}

\usepackage{graphicx}
\usepackage[body={6.5in, 9in}, right=1in, top=1in]{geometry}
\usepackage{amssymb}
\usepackage{amsmath}
\usepackage{amsthm}

\newcommand\ee{\end{equation}}
\newcommand\be{\begin{equation}}
\newcommand\eea{\end{eqnarray}}
\newcommand\bea{\begin{eqnarray}}

\newcommand\di{\partial}
\newcommand\mpl{M_{\rm Pl}}
\newtheorem{mydef}{Definition}

\begin{document}

\begin{center}
\LARGE{\textbf{Classical Stability of the Galileon}} \\[1cm]
\large{Solomon Endlich, and Junpu Wang}
\\[0.4cm]

\vspace{.2cm}
\small{\textit{ Department of Physics and ISCAP, \\
Columbia University, New York, NY 10027, USA}}

\end{center}

\vspace{.2cm}

\begin{abstract}
We consider the classical equations of motion for a single Galileon field with generic parameters in the presence of non-relativistic sources. We introduce the concept of absolute stability of a theory: if one can show that a field at a single point---like infinity for instance---in spacetime is stable, then stability of the field over the rest of spacetime is guaranteed for any positive energy source configuration. The Dvali-Gabadadze-Porrati (DGP) model is stable in this manner, and previous studies of spherically symmetric solutions suggest that certain classes of the single field Galileon (of which the DGP model is a subclass) may have this property as well. We find, however, that when general solutions are considered this is {\em not} the case. In fact, when considering generic solutions there are no choices of free parameters in the Galileon theory that will lead to absolute stability {\em except} the DGP choice. Our analysis indicates that the DGP model is an exceptional choice among the large class of possible single field Galileon theories. This implies that if general solutions (non-spherically symmetric) exist they may be unstable. Given astrophysical motivation for the Galileon, further investigation into these unstable solutions may prove fruitful.
\end{abstract}

%%%%%%%%%%%%%%%%%%%%%%%%%%%%%%%%%%
%%%%%%%%%%%%%%%%%%%%%%%%%%%%%%%%%%
\section{Introduction}

\subsection{The Galileon}

The Galileon naturally arises when one considers ghost free theories that modify General Relativity in the IR \cite{Galileon} (see also, \cite{Galileon Paper 1}-\cite{Galileon Paper 6} and its covariant extension \cite{Cov 1} and \cite{Cov 2}, and its use as an alternative to inflation \cite{Galileon Genesis}). In a local patch smaller than the cosmological horizon the additional degrees of freedom that encode our new dynamics are given by a relativistic scalar field, denoted by $\pi$, which is universally coupled to matter via $\pi T^\mu_\mu$, where $\pi$ generically decouples from matter at short scales due to derivative self-interactions (dubbed the Vainshtein effect). The Lagrangian describing these extra degrees of freedom is given by
\be \label{Galileon Lag}
\mathcal{L}_\pi=\sum_{i=1}^5 M_{pl}^2 c_i \mathcal{L}_i \, , \quad \text{where schematically} \quad \mathcal{L}_n=(\di^2 \pi)^{n-2}\di \pi \di \pi
\ee
and the equation of motion for the field is
\be \label{eom}
\sum_{i=1}^5  c_i \mathcal{E}_i=-\frac{T^\mu\, _\mu}{M_{pl}^2} \, , \quad \text{where} \quad \mathcal{E}_n=\frac{\delta \mathcal{L}_n}{\delta \pi}=(\di \di \pi)^{n-1}
\ee

We arrive at this particular equation of motion by imposing the symmetry $\pi(x)\rightarrow \pi(x)+b_\mu x^\mu+c$. This symmetry will require the equations of motion to have {\it at least} two derivatives. However, a well defined Cauchy problem/absence of ghosts requires that the equation of motion have{ \it at most} two derivatives. These requirements are strong enough to uniquely determine each $\mathcal{L}_n$ (these can be found in \cite{Galileon}) to an overall constant, given here by the $c_i$'s. This structure is very powerful, as we can re-write the partial differential equation as algebraic one

\be
F(k_{\mu \nu})=-\rho\, , \quad \text{where }\: k_{\mu \nu }\equiv \di_\mu \di_\nu \pi \: \text{ and } \: \rho=T_\mu^\mu
\ee

In order for the Galileon to be an {\em interesting} modification to gravity (its original motivation), we demand that the non-linear nature of the Galileon theory becomes manifest at the cosmological scale; that is, we want the non-linearities to become important at Hubble scales. This allows, in principle, cosmological solutions driven by $\pi$'s self-interactions rather than by the usual matter $T_{\mu \nu}$. We rewrite $c_i \mathcal{L}_i \sim  \tilde c_i \di \pi \di \pi\left(\di^2 \pi/H_0^2\right)^{n-2}$ where the new dimensionless coefficients $\tilde c_i$, defined as the combination of original ones $c_i$ and Hubble constant $H_0$, should be of order unity. A quick word of clarification: when we make reference to the DGP model, we mean not strictly the coefficients that come from the 4-d effective field theory associated with the extra-dimensional model put forth by \cite{DGP-original}, but rather the {\em class} of theories where $\tilde{c}_1=0$, $\tilde{c}_2>0$, $\tilde{c}_4=0$, and $\tilde{c}_5=0$. Subsequently, Galileon terms have also appeared in the decoupling limit of an interesting interacting theory of Lorentz invariant massive gravity \cite{Massive Gravity} as well as in a supersymmetric extension of ghost condensate theories \cite{Supersymmetric Galileons}. Additionally, the following analysis could possibly be extended for the more general cases of multi-field Galileons \cite{MultiGalileons}.

In this paper, we are concerned with classical stability and the non-linearities perform a critical role. We can investigate such a classical question because we are assured that quantum effects are small. Thanks to the Galileon invariance, the $\mathcal{L}_n$'s above are not renormalized under loop corrections, and terms with fewer derivatives acting on the $\pi$ field are not generated quantum mechanically \cite{LPR} (a similar proof holds for the multi-field Galileons \cite{MultiGalileons}). In particular, upon estimating the one loop effective action one can see that for classical solutions with large non-linearities quantum effects are small provided gradients are mild $\di \ll \Lambda$ \cite{DGP}. All of the above indicates that it is consistent to focus on the stability of classical solutions to (\ref{eom}).

In \cite{Derrick} it was shown that there do not exist any stable, static, source free soliton solutions. Any such solution, if it existed, would be extremely unstable. This instability lies in `wrong sign' kinetic terms, and therefore instabilities exist all the way down to the UV cutoff. Such a solution, if you attempted to construct it, would simply exit your effective theory immediately. This should be contrasted to instabilities stemming from  `wrong sign' potential terms. In this case one can follow a constructed solution as it changes into something else while still maintaining perturbative control (like a ball slowly rolling down shallow hill). We ask the following: does the picture change for non-linear solutions when non-relativistic sources are added to self interactions? Are these classically stable? Some analysis has been done along these lines. In particular, radial solutions were throughly investigated in \cite{Galileon}. Given the recent interest in Galileon theories, we feel it prudent to build upon, generalize, and possibly clarify the work done for radial solutions.

This paper is organized as follows: In Section $2$, we specify our problem and describe the general logic of our analysis, followed by a brief report of our results. In Section 4, in order to make our results more transparent, some graphical and analytical interpretations are presented. Detailed (and lengthy) analysis is contained within Section 3 and the Appendix. Readers are invited to skip these sections as they have been designed to contain all the gory details. In Section 5, we summarize our findings and suggest avenues for future inquiry.

%%%%%%%%%%%%%%%%%%%%%%%%%%%%%%%%%%
%%%%%%%%%%%%%%%%%%%%%%%%%%%%%%%%%%
\section{Outline of Analysis}

We begin by assuming some solution, say $\pi_0$, for the Galileon field that satisfies the equations of motion in the presence of some source, and then expand around this solution by adding small fluctuations. Taking $\pi_0\rightarrow\pi_0+\phi$ and keeping second order in $\phi$ terms in the Lagrangian, we have the dynamics of the fluctuations.\footnote[1]{Throughout this paper we will be using the $(-,+,+,+)$ signature.} The action is given by
\be \label{fluctuation action}
S_\phi=\frac{1}{2} \int d^4x Z^{\mu \nu}(x) \di_\mu \phi \di_\nu \phi
\ee
where $Z_{\mu \nu}$ is a matrix made out of the second derivatives of our $\pi_0$ field (i.e. made out of the matrix $k_{\mu \nu}$).\footnote[2]{The reason why only terms proportional to $\di \phi \di \phi$ survive in (\ref{fluctuation action}) is as follows: Note that the $n$th order Lagrangian can be written as $\mathcal{L}_n = T^{\mu_1 \nu_1 \dots \mu_n\nu_n}\di_{\mu_1}\pi\di_{\nu_1}\pi\di_{\mu_2 \nu_2}\pi\dots\di_{\mu_n \nu_n}\pi$, where $T$ is antisymmetric under changes of any ($\mu_i, \mu_j$) or ($\nu_i, \nu_j$) pair while symmetric under that of ($\mu_i, \nu_i$) pair \cite{Galileon}. Thus, thanks to the properties $T$ possesses,  all the terms in the second variation of $\mathcal{L}_n$ can be expressed as $\di \phi \di \phi (\di \di \pi_0)^{n-2}+ \text{surface terms}$ }

On the other hand, one can show that the matrix $Z^{\mu \nu}$ can be found by taking the first order variation of the {\em equation of motion}
\begin{equation}
\delta \mathcal{E}_n = - Z^{\mu \nu} \di_\mu \di_\nu \phi
\end{equation}
which provides an easier way to compute $Z$.\footnote[3]{Note that this is consistent with (\ref{fluctuation action}) because $\di_\mu Z^{\mu \nu}=0$ identically.} Simply put, we will investigate under what conditions these small fluctuations remain small.
%%%%%%%
\subsection{Conditions for Stability}

The main goal of this paper is to explore the parameter space spanned by the $c_i$'s ($i=1\,\dots,5$) to determine if there exists a subspace in which the Galileon theory is stable. We should point out that the ``stability" we refer to is consistent with the physical meaning meant by \cite{DGP}. While there are other definitions of stability, we utilize this definition because, though it may be limited, it is precise. Given that it is slightly different than the usual definition of stability in ODEs or PDEs, it is worth a few explanations. First, the stability we consider here is a {\em local} one, i.e., on a space and time scale much shorter than those typical of the background field $\pi_0$, and thus we are safe to treat the matrix function $Z_{\mu \nu}(x)$ as constant. Therefore (it should be noted) we will not be able to keep track of phenomena like resonances which can be interpreted as ``instabilities''. These ``instabilities'' are of a much less catastrophic nature than the ones we discuss and, as previously mentioned, still allow for much analytic control. In the neighborhood of a given point, our stability corresponds to demanding the e.o.m. of the fluctuation field
\be
Z_{\mu \nu} \di^\mu \di^\nu \phi=0
\ee
give an oscillating solution. When cast in Fourier space, oscillating solutions correspond to

\be
(Z_{00}\omega + Z_{0 i}q^i)^2 = (Z_{0i}Z_{0j}-Z_{00}Z_{i j})q^i q^j
\ee
having real solutions for $\omega$ for any real spatial momentum vector $\vec{q}$, or equivalently, the matrix $Z_{0i}Z_{0j}-Z_{00}Z_{i j}$ will be positive definite. Second, we want the Galileon theory to be absent of ghost-like instability, i.e. the sign of the kinetic term of the fluctuation action (\ref{fluctuation action}) to be correct, which requires $Z_{00}>0$.

As previously mentioned, we will focus on non-relativistic matter sources, not only because of their great importance, but because it can be shown that, given such sources at a generic spacetime point, the symmetric tensor $k_{\mu \nu}$, and therefore $Z_{\mu \nu}$, can be diagonalized through a Lorentz transformation \cite{DGP}. Thus, the conditions for the local stability are simply
\be
Z_\mu<0 \label{localstability}
\ee
where $Z_\mu$'s are the diagonal elements of the matrix $Z^\mu \,_\nu \equiv \text{diag} (Z_0\,,Z_1\,,Z_2\,,Z_3)$.

It has been shown in the DGP model that, given positive energy density sources, if a specific solution is stable at some point (in the way indicated above) then its stability throughout the spacetime is assured \cite{DGP}. This is, of course, a desirable property of a theory. One may wonder whether a generic Galileon theory shares this same nice property. Are there ``safe'' choices of the $c_i$ parameters (evidently, they should include the DGP parametrization) such that this subclass of Galileon theories possess the same property as the DGP model? The answer to this question is the main result of this paper.

In order to be absolutely clear about what we accomplish, we define the concept of {\em absolute stability}. Assume some solution to the equations of motion, $\pi_0$, exists.

\begin{mydef}
An {\bf absolutely stable region} in parameter space is a region of $\{c_1,...,c_5\}$'s where, if at a single point in spacetime, say $x^\mu_0$, $Z_\mu(\pi_0(x^\mu_0))<0$ (i.e. $\pi_0$ is stable at this point), then for non-relativistic source profiles satisfying $\rho \in [0,\infty)$ the equations of motion guarantee that $Z_\mu<0$ over the rest of spacetime (i.e. $\pi_0$ is stable over all of space).
\end{mydef}

Why is this a useful concept? If a choice of parameters is absolutely stable, then it follows that for any non-relativistic positive energy source configuration---no matter what the global structure ---stability of a particular solution at a single point implies global stability. When talking about absolute stability one does not have to solve (or at least characterize the solutions) the Cauchy problem for all possible source configurations. We consider the whole equation of motion surface \footnote{When we formulate our analysis in terms of the eigenvalues of the $k_{\mu \nu}$ tensor, the equation of motion $\mathcal{E}=\rho/\mpl^2$ (at a single point in spacetime, or equivalently, with $\rho$ fixed) defines a surface in the space spanned by these eigenvalues. It is this surface that we are referring to. All the surfaces generated by different sources ($\rho\in [0,\infty)$ ) we group into a ``family".} rather than characterize particular solutions. This stronger cut on acceptable parameters allows us to side-step the difficulties of dealing with the arbitrarily complicated global structure of our source. Considering that we are dealing with a non-linear PDE it is surprising that we can say anything at all. In the general case we don't know how to show existence, but we are still able to say something about stability. \footnote{It should be noted that ``absolute stability'' does not imply that {\em all} solutions are stable, as there can be different branches in our equation of motion surface. But, once again, it does imply that if a solution is stable at one spacetime point it is stable over the rest of spacetime.}

The DGP theory is absolutely stable \cite{DGP}. Spherically symmetric solutions (and mild deformations of them) for particular choices of parameters in a single field Galileon model \cite{Galileon} are stable. Is there some part of this parameter space that admits absolute stability?

To our surprise, we find that the DGP model is the {\em single} absolutely stable class. That is, the powerful property of absolute stability that the classical DGP theory possesses does not carry over into the general Galileon theory.

%One resolution to avoid all these is as following: assuming that at a generic point $x^\mu$, each matrix element of $k_{\mu \nu}$ can take any (real) value and the density of the source $\rho$ any positive value, subject only to the e.o.m.(\ref{eom}), restrict the parameters $c_i$'s such that into a subspace in which for all possible $k_{\mu \nu}$ the local stability is guaranteed(i.e. inequality (\ref{localstability}) is satisfied). Obviously, by doing that we actually strengthen the stability requirement suggested above and therefore we would refer this parameter subspace as "absolutely stable region" (ASR). The task for the rest of this paper is simply to determine such region for the galileon theory. However we want to emphasize that even if the $c_i$'s take values outside the ASR, we can {\em not} assert that the galileon theory will yield unstable solution; after all, the procedure of finding the ASR is only a shortcut of analyzing the stability without solving the e.o.m..%

%%%%%%%%%%%%%%%%%%%%%%%%%%%%%%%%%%%%%%%%%%%%%
\subsection{General Program}

After diagonalizing $k_{\mu \nu}$ with an appropriate boost we can write the stability conditions in a nice algebraic way:
\bea
Z^0\,_0&\equiv& Z_0(c_2,c_3, c_4,c_5,k_1,k_2,k_3)<0\\
Z^1\,_1&\equiv&Z_1(c_2,c_3, c_4,k_2,k_3)<0\\
Z^2\,_2&\equiv&Z_2(c_2,c_3, c_4,k_1,k_3)<0\\
Z^3\,_3&\equiv&Z_3(c_2,c_3, c_4,k_1,k_2)<0
\eea
where the $c_i$'s are the coefficients that describe our freedom in choosing the exact Galileon Lagrangain (\ref{Galileon Lag}) and the $k$'s are the eigenvalues of $k_{\mu \nu}$ for non-relativistic sources. The reason $k_0$ ($\equiv k_{00}$ after the matrix has been diagnalized) does not appear in the expression above is because it is suppressed by two powers of $v \ll 1$ in comparison to the other eigenvalues, that is $k_{00} \sim v^2 k_{i j}$, which must be small by assumption in order to ensure diagonalization.   $Z_0$ is a cubic function of the $k$'s while the $Z_i$'s are quadratic.

Additionally, we have the equation of motion for $\pi_0$ which becomes an algebraic equation for the $k$'s. Note that $c_1$ and $\rho$ enter in the same manner and are easily combined when we consider a single point in spacetime. We have
\be
\mathcal{E}(c_1,c_2,c_3, c_4,c_5,k_1,k_2,k_3)=\frac{\rho}{\mpl^2}
\ee
$\mathcal{E}$ is cubic in the $k$'s.

A particular choice of $c_i$'s define a given Galileon theory. They are considered constant over all of spacetime while the values of $\rho(x)$ that characterize our source will have various profiles for different physical configurations. Consider a single point in this space, say $\vec x_1$.

In eigenvalue space (what we will call ``$k$'' space) the equations of motion generate a surface (or more correctly surfaces---branches---as our equation is a cubic polynomial) which depend on $\rho(\vec x)$. The $Z_\mu<0$ inequalities will define volumes in $k$ space; they are independent of $\rho(\vec x)$. Say we restrict ourselves to a particular surface generated by $\mathcal{E}=\rho/\mpl^2$.

The question is: given particular values of the $c_i$'s and $\rho(\vec x_1)$ are there some values of the $k$'s that lie on this surface but violate  $Z_0<0$ or $Z_1<0$ (it is enough to consider $Z_0$ and a single $Z_i$)? If this is the case, then either the entire surface is imbedded inside a region where at least one $Z_\mu>0$, or this surface intersects with the `marginality surfaces'---the surfaces generated by $Z_\mu=0$\footnote[1]{Just for emphasis: the space that all these surfaces live in is, of course, the $(k_1,k_2,k_3)$ space. The only function (at this point) of the real space, $(x,y,z)$, is to give us the single point $x_1$ whose purpose is to pick out a value of our source, $\rho(\vec x_1)$. }. If this particular surface of solutions fails to intersect with any of these marginal surfaces, and has at least one particular choice of $k$'s where the stability inequalities hold, then we say that this surface is a {\em stable surface} at the point $\vec x_1$.

Repeating this analysis at a different point in space, say $\vec x_2$, means only taking $\rho(\vec x_1)\rightarrow\rho(\vec x_2)$. Beyond restricting the sources to positive ones, $\rho(\vec{x})\ge0$, a priori we have no idea what the source profile, $\rho(\vec{x})$, will be. Thus, in order to ensure stability of a solution generated by a particular source configuration $\rho(\vec x)$, we want to find the $\{c_i\}'s$ such that the family of e.o.m. surfaces generated by the possible values of $\rho(\vec x)$ is fully embedded within the stable regions ($Z_\mu<0$). That is, the possible e.o.m. surface at any spatial point is a stable surface.

Usually, the e.o.m. surface $\mathcal{E}=\rho/\mpl^2$ will have multiple branches. Consider a particular solution $\pi_0$ that at one point  $\vec x_0$ in real space is a point (in $k$ space) sitting on a particular branch of the e.o.m. surface $\mathcal{E}=\rho(\vec x_0)/\mpl^2$. If our solution is continuous then at a different point $\vec x_0+ \delta \vec x$ this particular solution will on be the same branch. That is, the branch it is on now is obtained from the previous branch under a continuous change of $\rho: \rho(\vec x_0)\rightarrow \rho(\vec x_0+\delta \vec x)$. Solutions are confined to a single branch---they cannot jump from one to another. We can therefore analyze each branch in isolation.

For the purposes of our proof there are two marginal surfaces in $k$ space we will be concerned with. The first is defined by $Z_0=0$ and the second by $Z_1=0$. We will consider their intersections with the $\mathcal{E}=\rho/\mpl^2$ surface.  If it is possible that no intersection occurs, then we must check which side (the stable or unstable) the particular surface generated by the equations of motion falls. In total, consideration of each intersection will generate a set of `stable' (in the sense given above) $\{c_i\}'s$ for each marginal surface---if they exist. The intersection of both these sets will be the stable choices of the $\{c_i\}'s$ for that particular $\mathcal{E}=\rho/\mpl^2$ surface. The intersection of these stable choices for every positive $\rho$ will give the absolutely stable values of the $\{c_i\}'s$ for the single field Galileon.

In summary, searching for absolute stability corresponds to looking for values of the $c_i$'s (if any) where some particular branch of any e.o.m surface $\mathcal{E}=\rho/\mpl^2$ does not inhabit any of the volume excluded by the stability inequalities regardless of the value of $\rho$. Given that the entire e.o.m. surface steers clear of regions of instability, we are assured that if there exists a solution whose value at one point in $x$ space happens to be associated with some point on the stable branch of the e.o.m. surface in $k$ space then this solution is stable over all of space. Calculationally, we  determine the absolutely stable region by taking the following steps: 1) find the conditions for $\{c_i\}$'s such that there are no real solutions to the algebraic equations $Z_\mu(k_1,k_2,k_3)=0$ and $\mathcal{E}(k_1,k_2,k_3)=\rho/\mpl^2$ (for a fixed $\rho$), 2) check that some point on the $\mathcal{E}(k_1,k_2,k_3)=\rho/\mpl^2$ surface satisfies $Z_\mu<0$ and 3) repeat this process for all $\rho \in [0,+\infty)$ to obtain the intersection.

%%%%%%%%%%%%%%%%%%%%%%%%%%%%%%%%%%
%%%%%%%%%%%%%%%%%%%%%%%%%%%%%%%%%%
\section{Details of Analysis}

%Before we begin, there is an important constraint. When one considers the stability of {\it spherically symmetric} solutions, as was done in \cite{Galileon}, one has that $c_2>0$ and that $\frac{2c_3^2}{c_4}-3c_2>0$.

As previously mentioned, in the presence of non-relativistic sources we can diagonalize the matrix $\di_\alpha \di_\beta \pi_0$ at a point by an appropriate Lorentz transformation. We can then write the $Z$'s, (\ref{fluctuation action}), and the equation of motion of $\pi_0$, (\ref{eom}), as
\be
Z_\mu(\pi_0)=-\left[c_2+2c_3\left[\left(\sum_{\alpha=0}^3 k_\alpha\right)-k_\mu\right]-6c_4\left[\left(\frac{1}{2}\sum_{\alpha \ne \beta}^3 k_\alpha k_\beta\right)-\sum_{\alpha}^3 k_\alpha k_\mu\right]+24c_5\left[ \frac{k_0 k_1 k_2 k_3}{k_\mu}\right]\right]
\ee

\be
\mathcal{E}=c_1+c_2\sum_\alpha^3 k_\alpha+c_3 \sum_{\alpha \ne \beta}^3 k_\alpha k_\beta+c_4\sum_{\alpha \ne \beta \ne \gamma}^3k_\alpha k_\beta k_\gamma+c_5\sum_{\alpha \ne \beta \ne \gamma \ne \delta}^3k_\alpha k_\beta k_\gamma k_\delta-\frac{\rho}{M_{pl}^2}=0
\ee
where $\di_\alpha \di_\beta \pi_0=k_{\alpha \beta} =\text{diag} (k_0,k_1,k_2, k_3)$. As mentioned above, classical stability corresponds to $Z_0\, , Z_1\, , Z_2\, , Z_3<0$.

%%%%%%%%%%%%%%%%%%%%%%
%By our purpose, the parameters $\{c_i\}$ should be restricted to a region such that the e.o.m. surface, i.e. a $3D$ surface in the $4D$ $\{k_0\,,k_1\,,k_2\,,k_3\}$ space defined by the equations of motion of the the $\pi$ field
%\be
%\mathcal{E}=c_1+c_2\sum_\alpha^3 k_\alpha+c_3 \sum_{\alpha \ne \beta}^3 k_\alpha k_\beta+c_4\sum_{\alpha \ne \beta \ne \gamma}^3k_\alpha k_\beta k_\gamma+c_5\sum_{\alpha \ne \beta \ne \gamma \ne \delta}^3k_\alpha k_\beta k_\gamma k_\delta-\frac{\rho}{M_{pl}^2}=0
%\ee
%never intersects with any of the marginal surfaces, $Z_\mu=0$. If so, a solution found locally stable at some point is also stable at any other points in the universe.
%%%%%%%%%%%%%%%%%%%%%%%

Since we have a non-relativistic source we can consistently suppress the $k_0$ dependence in the above expressions as it is suppressed by $v^2$ and recover the static limit. In particular
\be
\mathcal{E}=c_1+c_2(k_1+k_2+k_3)+2c_3(k_1k_2+k_1k_3+k_2k_3)+6c_4(k_1k_2k_3)-\frac{\rho}{M_{pl}^2} \label{simeom}
\ee
and
\bea
Z_0&=&-\left[c_2+2c_3(k_1+k_2+k_3)+6c_4(k_1k_2+k_2k_3+k_1k_3)+24c_5k_1k_2k_3\right]\\
\label{Z1}
Z_1&=&-\left[c_2+2c_3(k_2+k_3)+6c_4(k_2k_3)\right]
\eea
and similarly for $Z_2$ and $Z_3$.

%For an given solution, suppose we start at some stable point, at which $Z_\mu<0$. In moving away from this point, generally (i.e.
%for a generic parametrization \{$c_i$\}) it may become unstable, with the signal being that the $3D$ surface in the $4D$ $\{k_0\,,k_1\,,k_2\,,k_3\}$
%space defined by eqn. (\ref{simeom}) intersects any of the marginal hyperplanes $Z_\mu$=0. Therefore, by our purpose, the parameters $\{c_i\}$ should be
%restricted to a region (hereinafter referred as stable choice) such that this could not happen. If so, a solution found locally stable at some point is also stable at any other points in the universe.

In the following subsections, we will proceed by analyzing the various possible scenarios---the $\mathcal{E}=\rho/\mpl^2$ surface intersecting with the $Z_0=0$ or $Z_1=0$ surfaces---independently, and then consider the intersection of their constraints. To claim absolute stability we must then further take the intersection of the combined constraints for all positive values of $\rho$. For the moment we will concentrate on cases with a nonvanishing $c_4$, and leave the various special cases associated with $c_4=0$ for the Appendix.

%\footnote{$c_4$ seems special in this regard, but this is only the case because we use $c_4$ (we could easily choose another dimensionful parameter like $c_3$) as our ``unit''. That is, in the following sections we create dimensionless parameters by multiplying the other parameters by appropriate powers of $c_4$.}

An important point that we prove in Appendix \ref{All Surfaces Intersect}: if one branch of the equation of motion surface intersects a marginal surface, then all others do as well. That is, it is enough to find {\em a single} intersection for a particular set of $c_i$'s to rule out absolute stability for that set.

\subsection{Conditions for intersection of the $\mathcal{E}=\rho/\mpl^2$ and $Z_0=0$ surfaces ($c_4 \ne 0$)}

Assume for the moment that we are in the stable region of the $Z_i$'s, that is $Z_1<0$, $Z_2<0$, and $Z_3<0$. We can express $k_1$, $k_2$, and $k_3$ in terms of the particular values of the $Z_i$'s. These are
\bea
\label{k's}
k_i=-\frac{c_3}{3c_4} \pm \frac{\sqrt{2f_1 f_2 f_3}}{6c_4^{1/2}f_i},\quad \text{where} \label{branches}\\
f_i\equiv f(Z_i)=-3Z_i+\frac{2c_3^2}{c_4}-3c_2 \geqq \frac{2c_3^2}{c_4}-3c_2\geqq 0
\eea
The last inequality comes from constraints obtained by analyzing the stability and existence of radial solutions of the Galileon theory \cite{Galileon}
\be
\begin{cases}\label{constraintspecial}
&c_2 >0 \\
&c_3 \ge \sqrt{\frac{3}{2}c_2 c_4} \\
&c_4 \ge 0 \\
&c_5 <0 \\
\end{cases}
\ee
% Note that the same analysis forced $c_4\ge0$ and $c_3\ge0$ {\bf DOUBLE CHECK THIS LAST STATEMENT--and maybe make more explicit where they come from (existence of solutions i think)}.

We normalize the field such that it doesn't carry any dimensions, that is $\left[ \pi \right]=M^0=1$ where the brackets mean the usual ``dimensions of'' and $M$ means "dimensions of mass". As the action is dimensionless in Planck units we have that
\be
\left[ c_{n}\right]=\left[M^2\right]^{2-n}
\ee
In order to compare the free parameters in the Lagrangian we define dimensionless quantities from the dimensionful $c_i$'s
\be
 \alpha_1\equiv \left( c_1-\frac{\rho}{M_{pl}^2}\right)\sqrt{c_4}, \quad \alpha_2\equiv c_2,  \quad \alpha_3\equiv\frac{c_3}{\sqrt{c_4}}, \quad \alpha_5\equiv \frac{c_5}{c_4^{3/2}}
\ee
From (\ref{constraintspecial}), it immediately follows that
\be
\alpha_3\ge\sqrt{\frac{3}{2}\alpha_2}\,, \quad \text{and } \alpha_5<0
\ee

Note that while originally we had the whole set of parameters $\{c_1, c_2, c_3, c_4, c_5 \}$, a stable choice of the these parameters depends only on the choice of the dimensionless parameters $\{\alpha_1, \alpha_2, \alpha_3, \alpha_5 \}$.
The $c_4$ dependence disappears because we use it as our units---we measure everything in terms of $c_4$. \footnote{We could have used any other dimensionfull  parameter, but we we find $c_4$ a convenient choice.} At the classical level, provided it has the correct sign, the overall normalization of the Lagrangian does not matter. Noting that  $\alpha_2=c_2>0$, we can therefore work with $\left\{ \frac{\alpha_1}{\alpha_2},\frac{\alpha_3}{\alpha_2},\frac{\alpha_5}{\alpha_2} \right\}$ which for brevity of notation we define as $\left\{ \alpha_1,\alpha_3,\alpha_5\right\} $. Equivalently, we are free to normalize our Lagrangian with the simple choice of $c_2=\alpha_2=1$. Either way, we are left with three parameters $\{ \alpha_1, \alpha_3, \alpha_5 \}$. We are now ready to investigate the intersection conditions.

Inserting our solutions for the $k_i$'s, (\ref{k's}), into $Z_0$ and the equation of motion, we have
\bea
\label{cond 1}
Z_0=-c_2+\frac{8\alpha_3^3\alpha_5}{9}-\left(\frac{1}{3}-\frac{4\alpha_3\alpha_5}{9} \right)t \mp \left(\frac{2\sqrt{2}\alpha_5}{9} \right)\delta \mp \left(-\frac{\sqrt{2}\alpha_3}{3}+\frac{4\sqrt{2}\alpha_3^2\alpha_5}{9} \right)u\delta=0\\
\label{cond 2}
\mathcal{E}=\alpha_1-\alpha_3+\frac{4\alpha_3^3}{9} \pm \frac{\delta}{9\sqrt{2}} \pm \left(\frac{1}{3\sqrt{2}}-\frac{\sqrt{2}\alpha_3^2}{9}\right)u\delta=0
\eea
where
\be
t=f_1+f_2+f_3, \quad u=\frac{1}{f_1}+\frac{1}{f_2}+\frac{1}{f_3}, \quad \delta=\sqrt{f_1 f_2 f_3}
\ee

Using (\ref{cond 1}) and (\ref{cond 2}) we can always express $u$ and $\delta$ in terms of $t$, which we will treat as a free parameter. That is, for some fixed value of $t$ we can solve for $u(t)$ and $\delta(t)$ using the above constraints. Furthermore, we can solve for $f_1$, $f_2$, and $f_3$ via the algebraic equation
\be
F_t(x)\equiv x^3-tx^2+u(t)\delta(t)^2x-\delta(t)^2=(x-f_1)(x-f_2)(x-f_3)=0
\ee
Finally, we can then invert
\be
f(Z_i)=-3Z_i+2\alpha_3^2-3\alpha_2
\ee
to obtain the $Z_i$'s.

{ \bf `Instability'}: For a particular choice of the $c_i$'s or ($\alpha_i$'s) such that for some value of $t$, the $Z_i$'s are found to be negative implies that a solution could cross into the instability region. Thus, there could exist unstable solutions (solutions in the $Z_0>0$ volume). It is not guaranteed that the solution is unstable, but rather that this is a possibility. Using our terminology: there will be no absolute stability. Hence the quotes.

{\bf Stability}: A stable choice of $c_i$'s or ($\alpha_i$'s) corresponds to there being no intersection of the surface of equation of motion and the $Z_0>0$ volume of instability in the $\{k_1,k_2,k_3\}$ space. Thus, we are seeking a choice of $\alpha_i$'s such that, for any $t$, the equation $F_t(x)=0$ {\it cannot} have three real roots, all of which must be greater than $(2\alpha_3^2-3)$. This can be written in the statement

\begin{align}
\{t|  &t\ge 3\left(2\alpha_3^2-3 \right), \quad \Delta_3[F_t] \ge 0,  \quad F_t\left(2\alpha_3^2-3 \right) \le 0,\nonumber \\
&\text{and } F'_t\left( x \right) \ge 0 \quad \text{for any } x \le 2\alpha_3^2-3 \}=\emptyset  \nonumber
\end{align}
where $\Delta_3[F_t]$ is the discriminant of the cubic equation $F_t(x)$=0.

By virtue of the quadratic nature of $F'_t(x)$, the last condition can be further simplified. Indeed, the axis of symmetry of the upward-opened parabola $F'_t(x)=3x^2-2tx+u(t)\delta^2(t)$ is $x=\frac{t}{3} \ge 2\alpha_3^2-3$, so the last condition can be replaced by $F'_t\left(2\alpha_3^2-3\right)\ge0$.

Interestingly, one observes that both roots of solution \eqref{branches} yield the same value of $u(t)\delta^2(t)$ and $\delta^2(t)$ upon which our auxiliary function $F_t(x)$ depends. Therefore, both roots actually give the same condition for the stable choice of $\alpha$'s and henceforth we can focus on either one.

In summary, the stability condition coming from demanding  that the e.o.m. surface {\em not} intersect the $Z_0=0$ marginal plane reads
\be
\{t|  t\ge 3\left(2\alpha_3^2-3 \right),  \Delta_3[F_t] \ge 0,  F_t\left(2\alpha_3^2-3 \right) \le 0, \text{and }F'_t\left(2\alpha_3^2-3\right)\ge0 \}=\emptyset \label{Z0=0 conditions}
\ee

Now, as mentioned in our general outline, we still need to check that the surface generated by the e.o.m. is on the side of stability ($Z_0<0$) so that we know we are seeing absolute stability as opposed to guaranteed instability. But we hold off on this final check for just a moment.

%Now if $c_1\ne0$, we can simply replace
%To find $\{c_i \}$ that correspond to stable solutions, if they exist, we have to look for the intersection of
%In order to prove that a solution, if it exists, is stable we

\subsection{Conditions for intersection of the $\mathcal{E}=\rho/\mpl^2$ and $Z_1=0$ surfaces ($c_4 \ne 0$)}

Assume that we are in the $Z_2$, $Z_3$ and $Z_0$ stability region. Using the expression for $Z_1$ given by (\ref{Z1}) we are free to write the equation of motion as
\bea
\mathcal{E}=c_1+c_2(k_2+k_3)+2c_3(k_2k_3)-k_1Z_1-\frac{\rho}{M_{pl}^2}=0\\
\rightarrow \mathcal{E}=c_1+c_2(k_2+k_3)+2c_3(k_2k_3)-\frac{\rho}{M_{pl}^2}=0
\eea
Solving the above equation together with the $Z_1=0$ equation yields the solutions
\be
k_{2,3}=\frac{-3\alpha_1+\alpha_3 \pm \sqrt{\Gamma}}{6-4\alpha_3^2}\frac{1}{\sqrt{c_4}}\label{ksz_1=0}
\ee
where we have already bothered to normalize everything such that $\alpha_2=1$ and where $\Gamma=9\alpha_1^2+6-18\alpha_1 \alpha_3-3\alpha_3^2+8\alpha_1 \alpha_3^3$. The plus or minus indicated above means that $k_2$ must take the plus while $k_3$ must take the minus, or vice versa. We pick one. Plugging these solutions into the expressions for $Z_2$, $Z_3$ and $Z_0$, all of which are negative given our assumptions, yields
\bea
Z_0&=&-\left(\frac{2 k_1 \sqrt{c_4}}{2\alpha_3^2-3}\right)(9\alpha_1-6\alpha_3+2\alpha_3^3+6\alpha_5-12\alpha_1 \alpha_3 \alpha_5)\le0\\
Z_2&=&-\frac{(9\alpha_1-9\alpha_3+4\alpha_3^3)k_1\sqrt{c_4}+(-3+3\alpha_1\alpha_3+\alpha_3^2)+
\sqrt{\Gamma}(3k_1\sqrt{c_4}+\alpha_3)}{2\alpha_3^2-3} \le0\\
Z_3&=&-\frac{(9\alpha_1-9\alpha_3+4\alpha_3^3)k_1\sqrt{c_4}+(-3+3\alpha_1\alpha_3+\alpha_3^2)-
\sqrt{\Gamma}(3k_1\sqrt{c_4}+\alpha_3)}{2\alpha_3^2-3} \le0
\eea
The latter two conditions imply
\be
(9\alpha_1-9\alpha_3+4\alpha_3^3)k_1\sqrt{c_4}+(-3+3\alpha_1\alpha_3+\alpha_3^2)\ge\sqrt{\Gamma}|(3k_1\sqrt{c_4}+\alpha_3)|
\ee

A stable choice of parameters corresponds to
\begin{multline}
\label{Z1=0 cond full}
\{k_1\sqrt{c_4}\;|\;k_1\sqrt{c_4}(9\alpha_1-6\alpha_3+2\alpha_3^3+6\alpha_5-12\alpha_1 \alpha_3 \alpha_5)\ge 0, \; \Gamma \ge0, \\
\text{ and } \left((9\alpha_1-9\alpha_3+4\alpha_3^3)k_1\sqrt{c_4}+(-3+3\alpha_1\alpha_3+\alpha_3^2)\right)\ge\sqrt{\Gamma}|(3k_1\sqrt{c_4}+\alpha_3)|  \}=\emptyset
\end{multline}
which could be further simplified upon taking into consideration some of the information coming from (\ref{Z0=0 conditions}) to
\be
\alpha_3-\frac{4}{9}\alpha_3^3-\frac{\sqrt{2}}{9}(2\alpha_3^2-3)^{3/2}<\alpha_1<\alpha_3-\frac{4}{9}\alpha_3^3
+\frac{\sqrt{2}}{9}(2\alpha_3^2-3)^{3/2}\label{Z1=0 conditions2}
\ee
We will leave the lengthy algebraic analysis that generates the above conditions to Appendix \ref{AppendixB}, to which careful readers are referred.

\subsection{Stable Region: Local stability to absolute stability}\label{secsum}

We must, of course, take the intersection of the stable choices coming from both conditions, (\ref{Z0=0 conditions}) and (\ref{Z1=0 conditions2}).

%Doing this will also assure us that we don't have to worry about what happens when we hit multiple marginality surfaces at once with
%the equation of motion surface.
So far, we have worked out the stability conditions for a given choice of $\alpha_i$'s, with the external source $\rho$ being taken as a fixed parameter like the intrinsic ones ($c_i$'s) that define the theory. Say that, at the particular spatial point we are working at, the value of $\rho$ is such that the e.o.m. surface $\mathcal{E}=\rho/\mpl^2$ are completely imbedded in the stability region. As we move to a new point in space $\rho$ will generically change. Thus, we need to do our analysis all over again for this new value of the source. The convenience of our method is the details of how the source changes are washed out as we consider the whole family of surfaces generated by the e.o.m. rather than any particular solution. We lose some information, but we have made the problem tractable. We don't have to deal with the functional dependence of our source, $\rho(\vec x)$, boundary conditions, etc. All that matters is the range of values $\rho$ takes, $[\rho_{min},\rho_{max}]$. The values of $\rho$ encountered in any kind of astrophysical/cosmological application of the Galileon theory will be vast, spanning over 40 orders of magnitude from the average density of the universe ($\sim (10^{-3} eV)^4$) to nuclear density ($\sim (GeV)^4$). To describe a universe like our own in our units we can be free to take $\rho_{min}\rightarrow 0$ and $\rho_{max}\rightarrow \infty$.

% Since we don't know where, in real space, occurs the intersection of the e.o.m. surface with the $Z_\mu=0$ marginal hyperplane, and notice the factor that throughout the universe, the density could significantly varies, it is natural to demand that the conditions for stable choice (\ref{z0=0 conditions}) and (\ref{Z1=0 conditions}) to hold for any positive $\rho$.

%and thus have really only showed the stability at a single point, because $\alpha_1$ contains $\rho$ on top of $c_1$.
%When we move from point to point (in real space) the only parameter that can change is $\alpha_1$ as it contains our source. For a
%particular value of the source we are given a particular area in $\alpha_3$ and $\alpha_5$ space where solutions are stable. Let's call
%this set of stable parameter choices $A(\alpha_1)$. For the classical solution to be stable {\it over all of real space}  for all
%possible source configurations we need to look for the intersection of all the sets of different possible $\rho$'s. Thus:
For a particular value of $\rho$, we are given a particular volume in parameter space $\{c_1,c_3,c_5\}$ (in units of $c_4$) by satisfying (\ref{Z0=0 conditions}) and (\ref{Z1=0 conditions2}). Let's call this set of stable parameter choices $A(\rho)$. In order to achieve absolute stability we need to take the intersection of the $A(\rho)$'s of all possible $\rho$'s. Thus
\begin{align}
 \text{Absolutely stable region of } \{c_1,c_3,c_5\}'s= \bigcap_{\rho=0}^\infty A(\rho)
\end{align}

However, it is easily seen that (\ref{Z1=0 conditions2}) {\em cannot} hold for any source, since as $\rho \to \infty$, $\alpha_1$ becomes more and more negative (for a given $c_1$) and eventually fails to fall into the region specified by the fixed value of $\alpha_3$ (\ref{Z1=0 conditions2}). This validates us not checking that the surface generated by the e.o.m. lies in the stable region. In summary: From the analysis above, we see that for generic sources, no matter how we choose the parameters that define our theory (finite values of $c_1$, $c_3$, and $c_5$) the surface generated by the e.o.m. will  pierce the marginal surface generated by $Z_1=0$ soiling any hope of absolute stability in our theory.

Moreover, when examining the various special situations with $c_4=0$ discussed in Appendix, we find that there is {\em no} absolutely stable choice of parameters except for the $c_4=0,c_5=0$ case, which is exactly the DGP parameterization.

%%%%%%%%%%%%%%%%%%%%%%%%%%%%%%%%%%%%%%%%%%%
%%%%%%%%%%%%%%%%%%%%%%%%%%%%%%%%%%%%%%%%%%%

\section{Analysis of stability regions}

\begin{figure}
\begin{center}
\includegraphics[bb=6 0 569 181,width=1.00\textwidth]{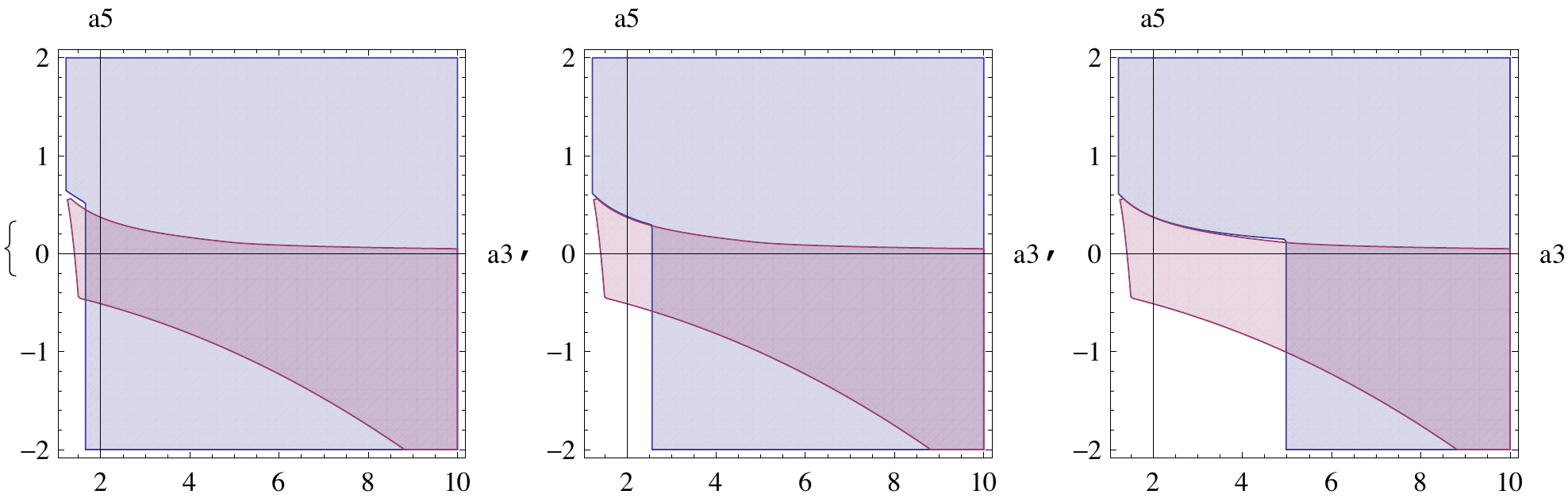}
\end{center}
\caption{Intersection of stability conditions given by $Z_0=0$ and $Z_1=0$ for (from left to right) $\alpha_1=-1$, $\alpha_1=-10$, and $\alpha_1=-100$. Red corresponds to stable regions for $Z_0=0$ and blue corresponds to stable regions for $Z_1=0$. The union of the two regions is the set of true stable choices for $\alpha_3$ and $\alpha_5$. }
\label{Z0=0 and Z1=0 stability regions}
\end{figure}

At this point the algebraic relations of (\ref{Z0=0 conditions}) and (\ref{Z1=0 conditions2}) don't provide much intuition. As an illustration we offer Figure \ref{Z0=0 and Z1=0 stability regions}, a plot of the stable regions associated with no intersection of the $Z_0=0$ and $Z_1=0$ marginal surfaces with the e.o.m surface for a particular choice of $\rho$.

The dynamics of the graph as a function of $\alpha_1$ are what we are interested in. There are obviously several curves that generate the boundaries that we see in Figure \ref{Z0=0 and Z1=0 stability regions}, but the one that is the most prominent is the vertical line forming the lower left boundary $Z_1=0$ region. To get an analytical sense of how the total region of stability depends on our source, we investigate the analytical behavior of this line. In particular, we see that it moves from left to right as we increase our source (decrease $\alpha_1$). We want to find out exactly how it moves as we vary the source. If, for instance, it moves logarithmically then it would not be effective at restricting the absolutely stable region (we would have to be more careful when we said that $\rho \rightarrow \infty$) whereas if it goes as some power law we can essentially claim that there is no absolutely stable region.

As shown in Appendix \ref{AppendixB} the key constraint in the $Z_1=0$ conditions is that some function (defined in the Appendix) $\Gamma(\alpha_1,\alpha_3)\ge0$. Setting $\Gamma(\alpha_1,\alpha_3)=0$ we find a particular solution $\alpha_3(\alpha_1)$. Examining this solution, we see that for large values of the source this line moves like
\be
\alpha_3 \simeq \left(-\frac{9\alpha_1}{8}\right)^{1/3}
\ee
and so we see that for generic sources there are no absolutely stable regions of parameter space (in $\alpha_3$ and $\alpha_5$) so that classical solutions of the equations of motion are guaranteed stability.

%%%%%%%%%%%%%%%%%%%%%%%%%%%%%%%%%%%%%%%%%%%
%%%%%%%%%%%%%%%%%%%%%%%%%%%%%%%%%%%%%%%%%%%

\section{Outlook and Conclusions}

After a careful analysis of all the various classes of Galileon theories (different values of the $c_i$'s), we find something fairly striking. It seems that {\em only} the DGP case is absolutely stable---the equations of motion don't intersect the marginality surfaces. Whereas for other, more general, Galileon theories the equations of motion {\em do} intersect the marginality surfaces.
This means that the strong statements given in \cite{Galileon} about regions of stability for spherically symmetric solutions does not extend to the general case. When generic solutions are taken into account the only absolutely stable region shrinks to that of the DGP model ($c_2>0$, $c_4=0$, and $c_5=0$). That is, the classical stability of the non-DGP Galileon theory as investigated by \cite{Galileon} is not as strong as suggested by that analysis. We want to emphasize that, while a lack of absolute stability does not guarantee instability (actual solutions could lie on the e.o.m. surface safely away from the marginal surfaces---the spherical solutions must do exactly this), it does seem to suggest that it could exist. When constructing (non-radial) solutions in generic Galillean theory, one needs to explicitly check classical stability via some method possibly valid only for those particular solutions. Of course, many physically interesting systems exhibit classical instabilities (gravitational collapse for instance). Our work suggests classical instabilities could be quite generic in single field Galileon theories. It would of great interest to find and investigate these potentially unstable solutions.

%%%%%%%%%%%%%%%%%%%%%%%%%%%%%%%%%%
%%%%%%%%%%%%%%%%%%%%%%%%%%%%%%%%%%
\section*{Aknowledgments}

We would like to thank Lam Hui, Kurt Hinterbichler, and especially Alberto Nicolis for illuminating discussions and guidance. We would also like to thank Katharine Lawrence for a careful editorial eye. The work of S.E. is supported by the National Science Foundation through a Graduate Research Fellowship. The work of J.W. is supported by the DOE (DE-FG02-92-ER40699).

%%%%%%%%%%%%%%%%%%%%%%%%%%%%%%%%%%%%%%%%%%%
%%%%%%%%%%%%%%%%%%%%%%%%%%%%%%%%%%%%%%%%%%%
\appendix

\section{Special Case of $c_4=0$}

\subsection{$c_5=0$, $c_3\ne0$}

This is the DGP-like case and has been thoroughly discussed in \cite{DGP}. For completeness, we quickly reproduce the arguments allowing for more general terms. Let's put $k_0$ back into our expressions, as it allows us to treat $Z_0$ and the $Z_i$'s in a similar manner for brevity (they all take a similar form). If any of the $Z_\mu=0$, for arguments sake $Z_0=0$, while the others remain $\le0$, then the equation of motion can be written
\be
\mathcal{E}=-\frac{c_2^2}{3c_3}+c_1-\frac{\rho}{M_{pl}^2}-\frac{1}{12c_3} \left\{(Z_1-Z_2)^2+(Z_1-Z_3)^2+(Z_2-Z_3)^2\right\}=0
\ee
Thus, if
\be -c_2^2+3c_3(c_1-\frac{\rho}{M_{pl}^2})<0 \label{DGPcase}
\ee
there will be no intersection. In particular, we want (\ref{DGPcase}) to hold for all $\rho\ge 0$. Therefore we demand that
\be
c_3>0,\,\quad \text{and } c_1<\frac{c_2^2}{3c_3} \label{DGPcase2}
\ee
Moreover, consider a solution with an known asymptotic behavior, (for instance, at infinity, it is deSitter-like: $k_0=k_1=k_2=k_3=k$). For a localized source, it is easy to show that the solution
\be
k \to \frac{-c_2+\sqrt{c_2^2-3c_1c_3}}{6c_3} \quad \text{as  } x^{\mu} \to \infty
\ee
is stable, since $Z_\mu=-c_2-6c_3 k=-\sqrt{c_2^2-3c_1c_3}<0$.

In the case of a non-localized source, or more complicated asymptotic behavior, we may be not able to check its stability directly. However, by constraint (\ref{DGPcase2}) we have absolute stability and so, provided our solution is stable at one point in space, we guarantee stability throughout the whole universe.

%%%%%%

\subsection{$c_3=0$, $c_5\ne0$}

$Z_i=-c_2<0$, so the only possibility is that the equation of motion $\mathcal{E}=\rho/\mpl^2$ (for some fixed $\rho$) may intersect with the $Z_0=0$ surface. In that case we can rewrite the $Z_0=0$ and $\mathcal{E}=0$ equations as
\bea
k_2 k_3&=&-\frac{c_2}{24c_5k_1}\\
k_2+k_3&=&-\frac{1}{c_2}(c_1-\frac{\rho}{M_{pl}^2})-k_1
\eea
We can see that $k_2$ and $k_3$ are the roots of the equation (treating $k_1$ as a free parameter)
\be
x^2+\frac{1}{c_2}(c_1-\frac{\rho}{M_{pl}^2}+c_2 k_1)x-\frac{c_2}{24c_5k_1}=0
\ee
for any $c_2>0$ and $c_5<0$ (remembering that these are conditions demanded by the analysis of absolute stability of spherical solutions) we can always choose a $k_1$ such that the above expression has two real roots, and thus an intersection between the marginality surface $Z_0=0$ and the surface generated by the equations of motion (for any $\rho$) occurs. According to our logic, we cannot say such a choice of our parameters leads to an unstable system, but rather that there is no absolutely stable choice of $c_i$'s where $c_3=c_4=0$.

%%%%%%

\subsection{$c_3\ne0$, $c_5\ne0$}

\subsubsection{Conditions for $\mathcal{E}=\rho/\mpl^2$ surface intersecting with $Z_0=0$ surface}

We can rewrite the $Z_0 =0$ and the $\mathcal{E}=\rho/\mpl^2$  equations as
\bea
K_1 K_2 K_3&=&-\frac{1}{24\beta_5}(c_2+2t)\\
K_1 K_2+K_1 K_3 +K_2 K_3&=&-\frac{1}{2}(\beta_1+c_2t)
\eea
where we have defined the dimensionless parameters
\be
K_i=c_3k_i,\quad \beta_1=(c_1-\frac{\rho}{M_{pl}^2})c_3, \quad \beta_5=\frac{c_5}{c_3^3}, \quad t=K_1+K_2+K_3
\ee
Note that
\begin{equation*}
Z_1=-\left(c_2+2(K_2+K_3)\right),,\quad Z_2=-\left(c_2+2(K_1+K_3)\right),,\quad Z_3=-\left(c_2+2(K_1+K_2)\right)
\end{equation*}
It is easier to discuss a cubic equation (about $x$) whose three real roots are $K_1+K_2$, $K_1+K_3$, and $K_2+K_3$: $G_t(x)=0$, where
\bea
G_t(x)&=&(x-K_1-K_2)(x-K_1-K_3)(x-K_2-K_3) \nonumber\\
&=&x^3-2tx^2+[(K_1+K_2)(K_1+K_3)+(K_1+K_2)(K_2+K_3) \nonumber\\
& &+(K_1+K_3)(K_2+K_3)]x-(K_1+K_2)(K_1+K_3)(K_2+K_3)\nonumber\\
&=&x^3-2tx^2+[t^2 + (K1 K2 + K1 K3 + K2 K3)]x\nonumber\\
& &-(K_1 K_2 + K_1 K_3 + K_2 K_3) t + K_1 K_2 K_3\nonumber\\
&=&x^3-2tx^2+[t^2-\frac{1}{2}(\beta_1+c_2t) ]x+\frac{t}{2}(\beta_1+c_2t)-\frac{1}{24\beta_5}(c_2+2t)
\eea
where we used the identities
\begin{align*}
&(K_1 + K_2) (K_1 + K_3) (K_2 + K_3) = (K_1 K_2 + K_1 K_3 + K_2 K_3) (K_1 + K_2 + K_3) - K_1 K_2 K_3\\
\text{and }\\
&(K1 + K2) (K1 + K3)  + (K1 + K2) (K3 + K2) + (K1 + K3) (K3 + K2) \\
& =  (K1 + K2 + K3)^2 + (K1 K2 + K1 K3 + K2 K3)
\end{align*}

If the three real roots of $G_t(x)=0$ are all greater than $-c_2/2$ for some value $t\ge-3c_2/4$, in which case $Z_i<0$, then the equation of motion surface $\mathcal{E}=\rho/\mpl^2$ intersects the $Z_0=0$ surface. Thus, a stable choice of the $\beta_i$ parameters corresponds to
\begin{multline}
\left\{t\;\Big|\; t\ge-\frac{3c_2}{4},\; \Delta_3[G_t(x)]\ge0,\; G_t(-\frac{c_2}{2})\le0, \;   \text{ and } G_t'(x)\ge0 \text{ for any }x\le-\frac{c_2}{2}\right\}=\emptyset \label{c4zeroz=0}
\end{multline}
By the same argument as above, the last condition can be replaced by $G_t'(-\frac{c_2}{2})\ge0$.

%Let's look at what these conditions imply. Setting $c_2=1$ as we are free to do.
%\begin{description}
%\item{$ t\ge-\frac{3c_2}{4}$:} $\Leftrightarrow$ $ t\ge-\frac{3}{4}$
%\item{ $ \Delta_3[G(t)]\ge0$:} $\Leftrightarrow$ a complicated polynomial inequality about $t$
%\item{$G(-1/2)<0$:}  $\Leftrightarrow$ $(1+2t)\left[(6t+6\beta_1-3)\beta_5-1\right]\ge0$ as $\beta_5<0$
%\item{$G'(x)\ge0$: }$\Leftrightarrow$ $\frac{\di G(x,t)}{\di x}=3x^2-4tx+t^2-t-\beta_1/2$. The symmetric axis of this function is about $x=2t/3\ge-1/2$, so when $x\le-1/2$ $\di G/\di x$ is a monotonically decreasing function of x, and therefore $\frac{\di G(-1/2,t)}{\di x}\ge0$ $\Leftrightarrow$ $t^2+t+3/4-\beta_1/2\ge0$
%\end{description}

\subsubsection{Conditions for $\mathcal{E}=\rho/\mpl^2$ surface intersects with $Z_1=0$ surface}

Proceeding in a similar manner as above, we can use the same dimensionless parameters and the $Z_1=0$ and $\mathcal{E}=\rho/\mpl^2$ equations to write
\bea
K_2+K_3&=&-\frac{c_2}{2}\nonumber\\
K_2 K_3&=&\frac{1}{4}(c_2^2-2\beta_1)\label{c4zeroz_1=0}
\eea
Note that the $K_1$ dependence drops out of the $\mathcal{E}=\rho/\mpl^2$ equation, as the part that is proportional to $K_1$ is also proportional to $Z_1$, which by assumption equals zero. Solving equations (\ref{c4zeroz_1=0}), we get

\bea
K_2&=&\frac{1}{4}(-c_2-\sqrt{-3c_2^2+8\beta_1}) \nonumber\\
K_3&=&\frac{1}{4}(-c_2+\sqrt{-3c_2^2+8\beta_1})
\eea
and thus we can input these values into the expressions for $Z_0$, $Z_2$ and $Z_3$. They are simply
\bea
Z_0&=&-2K_1(1+3\beta_5\left(c_2^2-2\beta_1)\right) \nonumber\\
Z_2&=&-\frac{1}{2}\left(c_2+4K_1+\sqrt{-3c_2^2+8\beta_1}\right)\nonumber\\
Z_3&=&-\frac{1}{2}\left(c_2+4K_1-\sqrt{-3c_2^2+8\beta_1}\right)
\eea
If there exists any $K_1$ such that all these $Z$'s are negative (or vanish), the e.o.m. surface may intersect with the $Z_1=0$ hyperplane.  A stable choice of $\beta_i$'s thus corresponds to
\be
\{K_1| \;Z_0(K_1)\le0, \;Z_2(K_1)\le0, \;Z_3(K_1)\le0, \text{ and }-3c_2^2+8\beta_1\ge0\}=\emptyset \label{c4zerocond}
\ee

%We see immediately, that $\beta_1<3/8$ violates these inequalities, in particular the last one. Physically, as a large source
%corresponds to a large negative value of $\beta_1$ those values don't allow the e.o.m. to intersect with the $Z_1=0$ plane, so we don't
%really gain anything from this analysis. To be sure we have been complete let's additionally investigate the $\beta_1\ge3/8$ region.

Note that $Z_3 \le 0$ is sufficient for $Z_2 \le 0$. After normalizing $c_2 \text{ to } 1$, we find that
\begin{itemize}
 \item{If $\beta_1<\frac{3}{8}$, this is a stable region, since it violates the last inequality in (\ref{c4zerocond}). }
 \item{If $\frac{3}{8}\le \beta_1 <\frac{1}{2}$, the $Z_3\le0$ condition reads $K_1 \ge \sqrt{8 \beta_1-3}-1$, which nevertheless allows $K_1$ to be either positive or negative, i.e. there exists some $K_1$ such that $Z_0<0$ regardless of the sign of $1+3\beta_5(1-2\beta_1)$. Therefore this is not a stable region.}
 \item{If $\beta_1\ge\frac{1}{2}$, $1+3\beta_5(1-2\beta_1)>0$ as $\beta_5<0$. For a sufficiently large $K_1>0$, all the inequalities in (\ref{c4zerocond}) are easily satisfied and so this is not a stable region.}
\end{itemize}

In summary, in order to avoid the intersection of the e.o.m. surface with the $Z_1=0$ marginal hyperplane, we should demand $\beta_1<3/8$.

%\begin{description}
%\item{$Z_3\ge0$:} $\Leftrightarrow$ $k_1\ge \sqrt{8\beta_1-3}-1$
%\item{$Z_0\ge0$:} $\Leftrightarrow$ $k_1\left[ (1+3\beta_5(1-2\beta_1)\right]\ge0$
%\end{description}

%Now, as $\beta_5<0$ for $\beta_1\ge1/2$ the inequalities are easily satisfied and so $\beta_1 \in [1/2,\infty)$ is not a stable region.
%Similarly, if $\beta_1 \in [3/8,1/2)$ the $Z_3\ge0$ condition allows $k_1$ to be either positive or negative which allows us to trivial
%find some $k_1$ that satsifies the $Z_0\ge0$ condition making this region also not a stable one. In summary, in order to ensure the
%e.o.m. surface never intersects the $Z_1=0$ plane $\beta_1<3/8$.

\subsubsection{Combining the conditions for the $\mathcal{E}=\rho/\mpl^2$ surface not to intersect with $Z_1=0$ and $Z_0=0$ surfaces}

If we restrict to $\beta_1<3/8$ we see that the $G_t'(-1/2)=(t+\frac{3}{4})^2+\frac{1}{2}(\frac{3}{8}-\beta_1)>0$ for any $t$, and so the last inequality in (\ref{c4zeroz=0}) gives us no new information. Thus, for $\mathcal{E}=\rho/\mpl^2$ to be a stable surface implies
\begin{align}
&\beta_1<3/8\,,\quad  \beta_5<0 \label{combinedc4zero1}\\
\text{and } \nonumber\\
&\left\{t \; \Big| \; \Delta_3[G_t(x)]\ge0, \; (1+2t)\left[(6\beta_1-3)\beta_5-1\right]\ge0,\; \text{and } t\ge-\frac{3}{4} \right\}= \emptyset \label{combinedc4zero2}
\end{align}
Now it is time to remind ourselves of the different roles of $c_1$ and $\rho$, both of which are wrapped up in the definition of $\beta_1$. Since the requirements of a stable region only constrain the intrinsic parameters (the $c_i$'s), we demand that (\ref{combinedc4zero1}) and (\ref{combinedc4zero2}) hold for any positive $\rho$. As we will now show, this is not possible for a sufficiently large density (i.e. as $\beta_1 \to -\infty$).

Fix $\beta_5 \ne 0$, for large negative values of $\beta_1$, the second and third inequality in (\ref{combinedc4zero2}) gives $t\ge-\frac{1}{2}$. Note in particular that $t=\sqrt{-2\beta_1}>0$ satisfies this inequality. Now, consider
\be
\Delta_3[G_t(x)] \big|_{t=\sqrt{-2\beta_1}} =\frac{1}{8}(-2 \beta_1)^{5/2}\dots >0
\ee
where ``$\dots$'' indicates the subdominant terms for sufficiently negative $\beta_1$, i.e. we have found some $t$ satisfying all the constraints in (\ref{combinedc4zero2}), which implies that for any nonvanishing $\beta_5$ and a generic source, no choice of parameters $\{c_i\}$ is absolutely stable.

Interestingly, for a vanishing $\beta_5$ ($c_4=0$ as well), we recover the DGP-like case, which {\em is} a stable choice for any positive sources. For a non-zero $\beta_5$, stability is only assured provided we restrict the source such that
\be
s(\beta_5)<\beta_1<3/8
\ee
where $s(\beta_5)$ is some complicated function which goes to $-\infty$ as $\beta_5\rightarrow 0^{-}$. It is in this sense that the DGP model is special. It is an asymptotic state in comparison to all other possible models generated by different points in the parameter space. One has to drive $\beta_5\rightarrow 0^{-}$ in order to ensure stability.

%%%%%%

\section{Further details of analysis }\label{AppendixB}

%We can see immediately from Figure \ref{Z0=0 and Z1=0 stability regions} that the vertical line generated by the $Z_1=0$ region determines our asymptotic behavior. As our source $\rho\rightarrow \infty$ does this line reach some asymptotic limit? Or will it always continue marching to the right such that for any choice of parameters in our Lagrangian there exists a value for the source that fails to guarantee stability?

We will simplify equation (\ref{Z1=0 cond full}) using some algebraic tricks.
In analyzing the conditions for the $Z_1=0$ region we begin with the second condition in (\ref{Z1=0 cond full})
\be
\Gamma=9\alpha_1^2+6-18\alpha_1 \alpha_3-3\alpha_3^2+8\alpha_1 \alpha_3^3\ge0
\ee
setting $\Gamma=0$ and solving for $\alpha_1$
\bea
\alpha_1=\alpha_3-\frac{4}{9}\alpha_3^3-\frac{\sqrt{2}}{9}(2\alpha_3^2-3)^{3/2}\equiv \alpha_{-}\\
\text{or }\quad \alpha_1=\alpha_3-\frac{4}{9}\alpha_3^3+\frac{\sqrt{2}}{9}(2\alpha_3^2-3)^{3/2}\equiv \alpha_{+}
\eea
Note, that since $2\alpha_3^2-3>0$ both of these solutions are real. Thus, we can rewrite the second condition as
\be
\label{gammacond}\Gamma\ge0 \Leftrightarrow \alpha_1\le \alpha_{-} \, \text{ or } \, \alpha_1\ge\alpha_{+}
\ee

Now, let's consider the third constraint in (\ref{Z1=0 cond full})
\bea
&&\left((9\alpha_1-9\alpha_3+4\alpha_3^3)k_1\sqrt{c_4}+(-3+3\alpha_1\alpha_3+\alpha_3^2)\right)\ge\sqrt{\Gamma}|(3k_1\sqrt{c_4}+\alpha_3)|\\
\label{cond1}&& \, \, \,   \Leftrightarrow \, \left((9\alpha_1-9\alpha_3+4\alpha_3^3)k_1\sqrt{c_4}+(-3+3\alpha_1\alpha_3+\alpha_3^2)\right)\ge 0 \text{  and  }\\
\label{cond2}&& \quad \quad  \left((9\alpha_1-9\alpha_3+4\alpha_3^3)k_1\sqrt{c_4}+(-3+3\alpha_1\alpha_3+\alpha_3^2)\right)^2\ge \Gamma(3k_1\sqrt{c_4}+\alpha_3)^2
\eea
Notice that when $\alpha_1\le \alpha_{-}$, the prefactor to $k_1$,  $(9\alpha_1-9\alpha_3+4\alpha_3^3)\; \le -\sqrt{2}(2\alpha_3^2-3)^{3/2} <0$. Similarly, when $\alpha_1\ge \alpha_{+}$ the prefactor $\ge+\sqrt{2}(2\alpha_3^2-3)^{3/2} >0$. Thus, we can combine the inequalities (\ref{cond1}) and (\ref{gammacond})  as
\be
\label{finalcond1}\alpha_1\le \alpha_{-}\,  \text{ and }\,k_1\sqrt{c_4} \le K(\alpha_1,\alpha_3) \quad \text{or } \quad \alpha_1\ge \alpha_{+}\, \text{ and }\, k_1\sqrt{c_4} \ge K(\alpha_1,\alpha_3)
\ee
where
\be
K(\alpha_1,\alpha_3)\equiv \frac{(-3+3\alpha_1\alpha_3+\alpha_3^2)}{(9\alpha_1-9\alpha_3+4\alpha_3^3)}=-\frac{\alpha_3}{3}+\frac{(2\alpha_3^2-3)^2}{3(9\alpha_1-9\alpha_3+4\alpha_3^3)}
\ee
Now, consider (\ref{cond2}) which can be rewritten as
\bea
&&(3-2\alpha_3^2)^2\left\{1-2\alpha_1\alpha_3+k_1\sqrt{c_4}(-6\alpha_1+2\alpha_3)+(k_1\sqrt{c_4})^2(4\alpha_3^2-6) \right\}\ge0 \quad \Leftrightarrow\\
&&1-2\alpha_1\alpha_3+k_1\sqrt{c_4}(-6\alpha_1+2\alpha_3)+(k_1\sqrt{c_4})^2(4\alpha_3^2-6)\ge 0 \quad \Leftrightarrow\\
&& k_1\le \frac{1}{\sqrt{c_4}}\frac{3\alpha_1-\alpha_3-\sqrt{\Gamma}}{4\alpha_3^2-6}= k_2(\alpha_1,\alpha_3)\,\,\text{or } \, \, k_1 \ge \frac{1}{\sqrt{c_4}} \frac{3\alpha_1-\alpha_3+\sqrt{\Gamma}}{4\alpha_3^2-6}=  k_3(\alpha_1,\alpha_3)\label{finalcond2}
\eea
In the last line, (\ref{ksz_1=0}) was used.
Let's compare (\ref{finalcond2}) with our previous results (\ref{finalcond1}). There are two regions we need to concern ourselves with: $\alpha_1\le \alpha_{-}$ and $\alpha_1\ge \alpha_{+}$. Amazingly, we find the combined result is simply
\be
\alpha_1\le \alpha_{-}\,  \text{ and }\,k_1\sqrt{c_4} \le \sqrt{c_4} k_2(\alpha_1,\alpha_3) \quad \text{or } \quad \alpha_1\ge \alpha_{+}\, \text{ and }\, k_1\sqrt{c_4} \ge \sqrt{c_4} k_3(\alpha_1,\alpha_3)\label{finalcond3}
\ee
due to the observation that
\begin{align}
\sqrt{c_4} k_2(\alpha_1,\alpha_3)\le \sqrt{c_4} k_2(\alpha_{-},\alpha_3)=K(\alpha_{-},\alpha_3)\le K(\alpha_1,\alpha_3) \quad \text{for }\alpha_1 \le \alpha_{-} \\
\sqrt{c_4} k_3(\alpha_1,\alpha_3)\ge \sqrt{c_4} k_3(\alpha_{+},\alpha_3)=K(\alpha_{+},\alpha_3)\ge K(\alpha_1,\alpha_3) \quad \text{for }\alpha_1 \ge \alpha_{+}
\end{align}
%%%%%
%{\bf I think this part may be greatly simplified. All we need to do is say that if we are on one side then $k_2<K$ and if we are on the other side $k_3>K$ and thus we can be more restrictive than what is in (\ref{finalcond1}), but let me put it as is for now....}

%%%%%

In order to have guaranteed  stability we need the conditions contained in (\ref{finalcond3}) and the first condition of (\ref{Z1=0 cond full}), the $\alpha_5$ dependent one, to yield the null set. Equivalently, explicitly separating the two regions, we may write
\begin{itemize}
  \item If $\alpha_1\le \alpha_{-}$, for any $k_1\sqrt{c_4}\le \sqrt{c_4}k_2(\alpha_1,\alpha_3)$,  $k_1\sqrt{c_4} f(\alpha_1,\alpha_3,\alpha_5)<0$
  \item If $\alpha_1\ge \alpha_{+}$, for any $k_1\sqrt{c_4}\ge \sqrt{c_4} k_3(\alpha_1,\alpha_3)$,  $k_1\sqrt{c_4} f(\alpha_1,\alpha_3,\alpha_5)<0$
\end{itemize}
Where we have defined $f(\alpha_1,\alpha_3,\alpha_5)\equiv (9-12 \alpha_3 \alpha_5)\alpha_1-6\alpha_3+2\alpha_3^3+6\alpha_5$. Let's investigate when these conditions are satisfied -- when we are guaranteed classical stability.

\begin{description}
\item{{\bf Instability for $\alpha_1\le \alpha_{-}$}}

As we are interested in the region where $\alpha_3> \sqrt{3/2}$ and $\alpha_5<0$, we see that $(9-12 \alpha_3 \alpha_5)>0$ and thus
\bea
&&f(\alpha_1,\alpha_3,\alpha_5)\le f(\alpha_{-},\alpha_3,\alpha_5)\\
&&=(2\alpha_3^2-3)\left\{-\alpha_3 -\sqrt{4\alpha_3^2-6}+\frac{\alpha_5}{3}(4\alpha_3\sqrt{4\alpha_3^2-6}+8\alpha_3^2-6) \right\}\\
&&<0\quad \text{for any $\alpha_5<0$ and $\alpha_3> \sqrt{3/2}$ }
\eea

So, for $\alpha_1 \in (-\infty, \alpha_{-}]$, $f(\alpha_1,\alpha_3,\alpha_5)<0$ and $k_2(\alpha_1,\alpha_3)<0$, from which we can see that there always exists some $k_1 \sqrt{c_4}$ such that $k_1\sqrt{c_4}f(\alpha_1,\alpha_3,\alpha_5)$ is positive. Thus, for $\alpha_1 \in (-\infty, \alpha_{-}]$, the condition (\ref{Z1=0 conditions}) is {\em not} a null set and the corresponding parameter choice is not a stable one.

\item{{\bf Instability for $\alpha_1\ge \alpha_{+}$}}

Note that for this region $k_3(\alpha_1,\alpha_3)$ is a monotonically increasing function of $\alpha_1$
\be
k_3(\alpha_1,\alpha_3)\ge k_3(\alpha_{+},\alpha_3)=\frac{1}{3} \left( -\alpha_3+\sqrt{\alpha_3^2-\frac{3}{2}} \right)
\ee
and that $k_3(\alpha_{+},\alpha_3)<0$. There exists an $\alpha_{*}$ such that $k_3(\alpha_{*},\alpha_3)=0$. In fact, $\alpha_{*}=1/2\alpha_3$.

If $\alpha_{+}\le \alpha_1\le \alpha_{*}$, then $k_3(\alpha_{+},\alpha_3)\le k_3(\alpha_{1},\alpha_3)\le k_3(\alpha_{*},\alpha_3)=0$ and therefore the $k_1\sqrt{c_4}$'s, satisfying $k_1\sqrt{c_4}\ge\sqrt{c_4} k_3(\alpha_{1},\alpha_3)$, could be either positive or negative and so be $k_1 \sqrt{c_4}f(\alpha_1,\alpha_3,\alpha_5)$, regardless of the sign of $f(\alpha_1,\alpha_3,
\alpha_5)$. Once again, the condition (\ref{Z1=0 conditions}) is {\em not} a null set and the region $\alpha_1 \in [\alpha_{+}, \alpha_{*}]$ is not stable.

If $\alpha_1\ge \alpha_{*}$, then $k_3(\alpha_1,\alpha_3)\ge k_3(\alpha_{*},\alpha_3)=0$. Since
\be
f(\alpha_1,\alpha_3,\alpha_5)\ge f(\alpha_{*},\alpha_3,\alpha_5) =\frac{(2\alpha_3^2-3)^2}{2\alpha_3} >0
\ee
so it follows that for any $k_1\sqrt{c_4}$ ( $\ge  \sqrt{c_4} k_3(\alpha_{1},\alpha_3))$, $k_1\sqrt{c_4}f(\alpha_1,\alpha_3,\alpha_5)$ is positive. By the same argument above, we see that $\alpha_1\ge \alpha_{*}$ is not a stable region.
\end{description}

In summary, after a long and torturous process we see that the only stable region (i.e. the choice of parameter making condition (\ref{Z1=0 cond full}) a null set)  is determined by $\Gamma<0$, or in other words
\be
\alpha_{-}=\alpha_3-\frac{4}{9}\alpha_3^3-\frac{\sqrt{2}}{9}(2\alpha_3^2-3)^{3/2}<\alpha_1<\alpha_3-\frac{4}{9}\alpha_3^3+\frac{\sqrt{2}}{9}(2\alpha_3^2-3)^{3/2}=\alpha_{+}
\ee

%This is not a particularly useful form, what we would like to do is invert these relations so give a value for $\alpha_1$ find what are the allowed values of $\alpha_3$. For generic sources, $\alpha_1\ll0$ and since $\alpha_{+}>0$ it is enough to invert $\alpha_{-}$ which yields, for large values of the source:
%\be
%\alpha_3 \simeq \left(-\frac{9\alpha_1}{8}\right)^{1/3}
%\ee

%%%%%%

\section{Intersection of all branches of equation of motion surface $\mathcal{E}=\rho/\mpl^2$ with marginal surfaces}\label{All Surfaces Intersect}

In this paper we have shown that for a fixed $\rho$, there are conditions on the parameters $\{c_1,c_2,c_3,c_4,c_5\}$, such that the surface generated by equation of motion $\mathcal{E}=\rho/\mpl^2$ will never intersect the marginal surfaces $Z_\mu=0$. However, when $\rho$ is allowed to vary from $0$ to $\infty$, these conditions generally cannot hold, i.e. no matter how we choose the $c_i$'s (with the exception of the DGP-like case), for some $\rho'$, the e.o.m. surface will eventually intersect at least one of the critical hyperplanes.

However, since that e.o.m. surface has disconnected branches, it is possible that only some branches of the e.o.m. surface intersect a marginal surface, while others are entirely buried within the stable region $Z_\mu<0$ for all positive $\rho$. If there is a chance that this is the case, then we cannot claim that any particular choice of parameters is not absolutely stable until we investigate each individual branch.

Here we rule out this possibility by showing that on each branch of the e.o.m. surface there are regions where at least one of the $Z_\mu$'s is positive, for some appropriate positive $\rho$ (with the exception of the DGP-like case).

\begin{figure}
\begin{center}
\resizebox{\linewidth}{!}{\includegraphics[bb=0 0 346 354,width=1.0\textwidth]{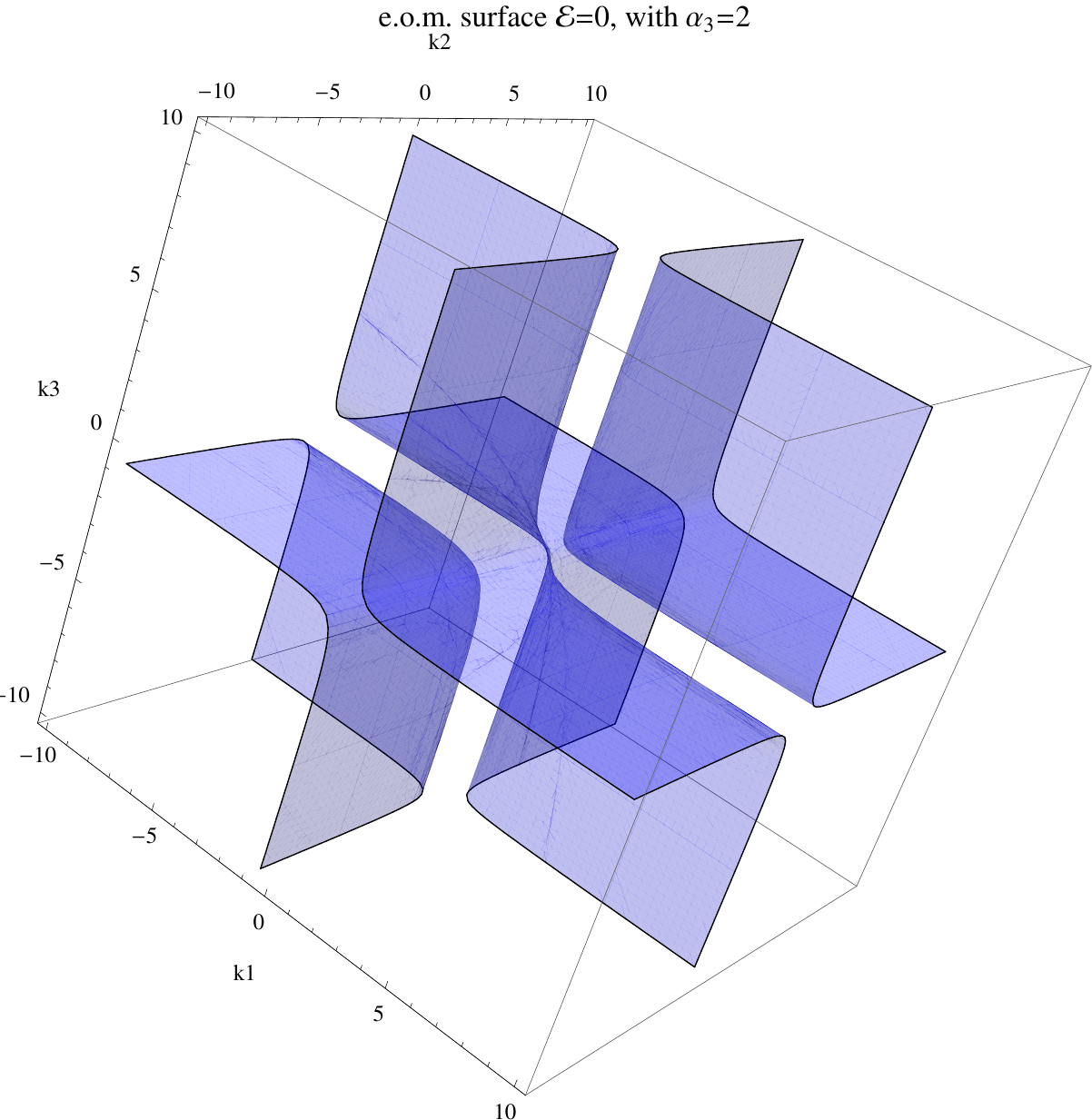}}
\end{center}
\caption{The surface generated by the equations of motion in `k' space with $\alpha_3=2$ and $\alpha_1=0$.}
\label{e.o.m. with brach 2 and 3 disconnected}
\end{figure}

Let's first focus on the case with $c_4\ne0$. With our usual notation and defining
\be
 K_i=\sqrt{c_4}k_i
\ee
The e.o.m. surface is given by
\be \label{AppendixEOMsurface}
\left\{(K_1,K_2,K_3)\big \vert K_1=-\frac{\alpha_1+K_2+K_3+2\alpha_3 K_2 K_3 }{6 K_2 K_3+2\alpha_3(K_2+K_3)+1} \right\}
\ee
Generally this surface insists of disconnected branches, due to the fact that for some $(K_2, K_3)$, the denominator of (\ref{AppendixEOMsurface}) vanishes. More precisely, suppose $(q_1,q_2,q_3)$ is some point on the e.o.m. surface (for any $\rho$), it is in (see Figure \ref{e.o.m. with brach 2 and 3 disconnected})
\begin{align*}
&\text{Branch 1, if } 6 q_2 q_3+2\alpha_3 (q_2+q_3)+1>0 \text{ and }q_2>-\frac{\alpha_3}{3}, q_3>-\frac{\alpha_3}{3}\\
&\text{Branch 2, if } 6 q_2 q_3+2\alpha_3 (q_2+q_3)+1<0\\
&\text{Branch 3, if } 6 q_2 q_3+2\alpha_3 (q_2+q_3)+1>0 \text{ and }q_2<-\frac{\alpha_3}{3}, q_3<-\frac{\alpha_3}{3}\\
\end{align*}

\begin{figure}
\begin{center}
\resizebox{\linewidth}{!}{\includegraphics[bb=21 0 284 275,width=1.0\textwidth]{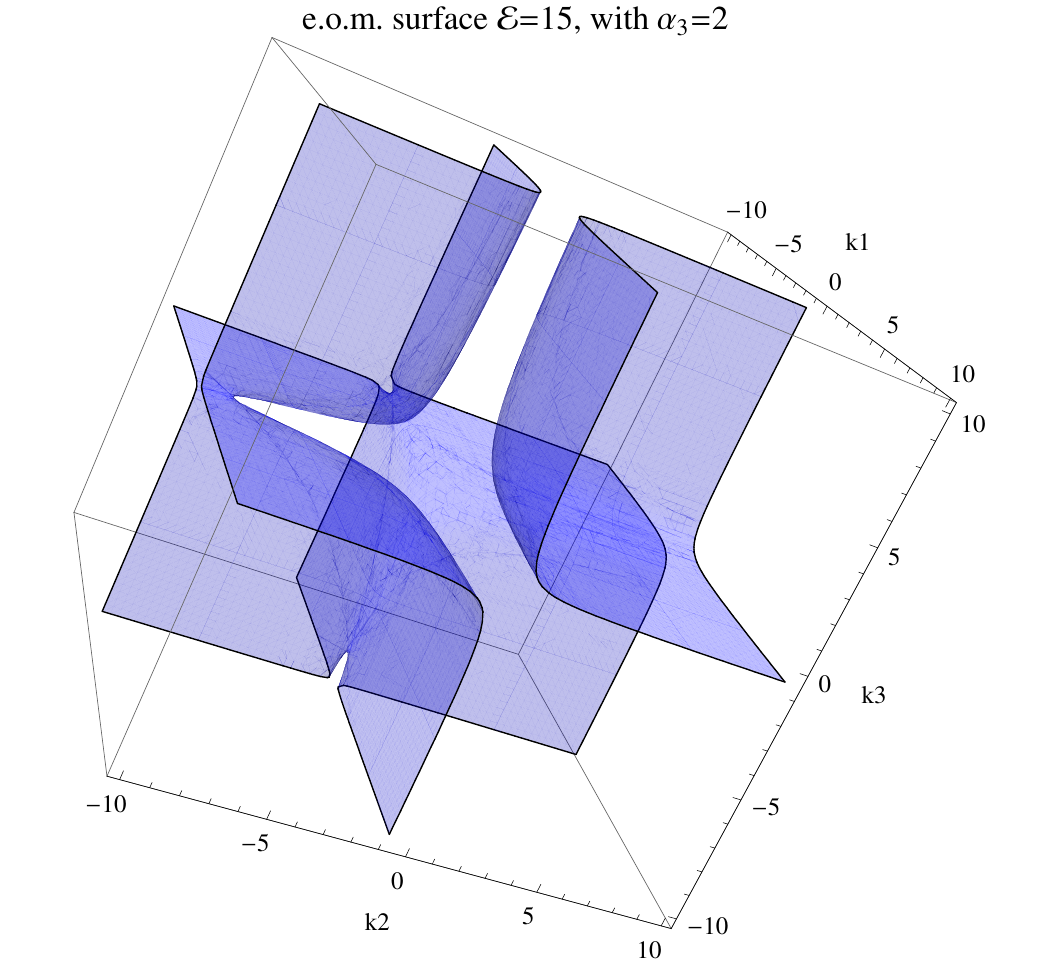}}
\end{center}
\caption{The surface generated by the equations of motion in `k' space with $\alpha_3=2$ and $\alpha_1=-15$.}
\label{e.o.m. with brach 2 and 3 connected}
\end{figure}

However, if there exists some $K_2,K_3$ satisfying simultaneously
\be
\begin{cases}
\alpha_1+K_2+K_3+2\alpha_3 K_2 K_3=0\label{connect}\\
6 K_2 K_3+2\alpha_3(K_2+K_3)+1=0
\end{cases}
\ee
some of the branches become connected (see Figure \ref{e.o.m. with brach 2 and 3 connected}).
 Indeed, (\ref{connect}) can be solved as
\be
\bar{K}_{2,3}=\frac{3\alpha_1-\alpha_3\pm\sqrt{\Gamma}}{4\alpha_3^2-6}
\ee
as long as $\Gamma=9\alpha_1^2+(-18\alpha_3+8\alpha_3^3)\alpha_1+6-3\alpha_3^2\ge0$. Equivalently
\be
\alpha_1\le\alpha_3-\frac{4}{9}\alpha_3^3-\frac{\sqrt{2}}{9}(2\alpha_3^2-3)^{3/2},\quad \text{or}\quad \alpha_1\ge\alpha_3-\frac{4}{9}\alpha_3^3
+\frac{\sqrt{2}}{9}(2\alpha_3^2-3)^{3/2}\label{Z1=0 conditions}
\ee
As $\rho\to\infty$, $\alpha_1\to-\infty$, and so for finite $\alpha_3$ it is easy to check that $\bar{K}_{2,3}<-\alpha_3/3$, i.e. in this limit, Branch 2 and Branch 3 become connected.

Pick some point $(K_1,K_2,K_3)$ on Branch 2 with
\be
K_2=K_3=-\alpha_3/3,\quad \text{and }K_1=\frac{9\alpha_1+2\alpha_3(\alpha_3^2-3)}{6\alpha_3^2-9}
\ee
Evaluating $Z_\mu$'s on this point, we find that $Z_1=(2\alpha_3^2-3)/3>0$; thus Branch 2 cannot be wholly contained within the stable region.

Similarly, consider some point on the e.o.m. surface, with $K_1=K_2=K_3=K$; in the limit $\rho\to \infty$ or $\alpha_1\to-\infty$, the equation of motion demands $K\simeq (-\alpha_1/6)^{1/3}$. It is straightforward to show that this point lies in Branch 1. Additionally, at this point $Z_0\simeq -4\alpha_5(-\alpha_1)>0$, i.e. Branch 1 can not be entirely within the stable region for all positive $\rho$'s.

Owing to the fact that Branch 2 and 3 will eventually join each other, we conclude (as promised) that none of the branches of the e.o.m. surface remain within the stable region $Z_\mu<0$ for all positive $\rho$'s. We can also run a similar argument for the $c_4=0$ (non-DGP-like) case and come to the same conclusion.

%%%%%%%%%%%%%%%%%%%%%%%%%%%%%%%%%%%%%%%%%%%
%%%%%%%%%%%%%%%%%%%%%%%%%%%%%%%%%%%%%%%%%%%

\end{document}